\documentclass[showpacs,amssymb,aps,twocolumn]{revtex4}
\usepackage{amsmath}
\usepackage{amstext}
\usepackage{amsopn}
\usepackage{amsfonts}
\usepackage{amssymb}
\usepackage{bbm}
\usepackage{accents}
\usepackage{empheq}
\usepackage{graphicx}
\usepackage{epsf}
\usepackage{graphics}
\usepackage[latin1]{inputenc}
\def\sc{\scriptscriptstyle}

\begin{document}

\title{Effective Actions at Finite Temperature}

\author{Ashok Das$^{a,b}$ and J. Frenkel$^{c}$\footnote{$\ $ e-mail: das@pas.rochester.edu,  
jfrenkel@fma.if.usp.br}}
\affiliation{$^a$ Department of Physics and Astronomy, University of Rochester, Rochester, 
NY 14627-0171, USA}
\affiliation{$^b$ Saha Institute of Nuclear Physics, 1/AF Bidhannagar, Calcutta 700064, India}
\affiliation{$^{c}$ Instituto de Física, Universidade de São Paulo, 05508-090, São Paulo, SP, BRAZIL}

\begin{abstract}
This is a more detailed version of our recent paper where we proposed, from first principles, a direct method for evaluating the exact fermion propagator in the 
presence of a general background field at finite temperature. This can, in turn,  be used to determine the 
finite temperature effective action for the system. As applications, we discuss the complete one 
loop finite temperature effective actions for $0+1$ dimensional QED as well as for the Schwinger 
model in detail. 
These effective actions, which are derived in the real time (closed time path) formalism, generate 
systematically all the Feynman amplitudes calculated in thermal perturbation theory and also show 
that the retarded (advanced)  amplitudes vanish in these theories. Various other aspects of the problem are also discussed in detail.
\end{abstract}


\maketitle
\newpage

\section{Introduction}

The effective action for a system of fermions interacting with a background field, which incorporates 
all the one loop corrections in the  theory, is an important fundamental concept in quantum field 
theory. At zero temperature we know that the $n$-point amplitudes (involving the background fields)  
at one loop are, in general,  divergent and, consequently, the evaluation of the effective action at 
$T=0$ needs a regularization. Effective actions can, of course, be evaluated perturbatively. 
However, a beautiful method due to Schwinger \cite{schwinger}, also known as the proper time 
formalism, is quite useful in evaluating one loop effective actions at zero temperature with a gauge 
invariant regularization (in the case of gauge backgrounds). We note that the effective action for  a 
fermion with mass $m$  interacting with a general background is given by
\begin{equation}
\Gamma_{\rm eff} [A] = - i {\rm Tr}\, \ln (i\partial - m - g A) = - i {\rm Tr}\, \ln H,\label{effaction}
\end{equation}
where $A$ denotes the background field and $g$ represents the coupling to the background and 
we have identified
\begin{equation}
H = i\partial - m - g A.
\end{equation}
Here we have suppressed the Lorentz structure 
of the kinetic term as well as the background to allow for generality. For example, the background 
field can be a scalar or a gauge background. Similarly, the system under consideration may be a 
($0+1$ dimensional) quantum mechanical system in which case the fermion kinetic term will have 
no structure (or contraction with Dirac gamma matrices).  

Schwinger expressed the effective action \eqref{effaction} in a regularized integral form
\begin{equation}
\Gamma_{\rm eff} [A] = \lim_{\nu\rightarrow 0}\ i \int\limits_{0}^{\infty} \frac{d\tau}{\tau^{1-\nu}}\,
{\rm Tr}\,e^{-\tau H},\label{effaction1}
\end{equation}
where $\tau$ is known as the ``proper time" parameter. The idea here is that the operator $e^{-\tau 
H}$ 
in the integrand can be thought of as the  evolution operator in the Euclidean time $\tau$ with $H$ 
denoting the (Hamiltonian) generator for the evolution. (The ``proper time" can also be made 
Minkowskian with appropriate $i\epsilon$ prescription.) As a result, we can write the proper time 
evolution equations
\begin{align}
\frac{dx^{\mu}}{d\tau}  & = -i [x^{\mu}, H],\notag\\
\frac{dp_{\mu}}{d\tau}  & = -i [p_{\mu}, H].\label{dynamicaleqn}
\end{align}
If these equations can be solved and $x^{\mu}(\tau)$ (or $p_{\mu} (\tau)$) can be determined in a 
closed form, then one can evaluate the trace in \eqref{effaction1} in the eigenbasis 
$|x^{\mu} (\tau)\rangle$ (or $|p_{\mu}(\tau)\rangle$) and evaluate the (gauge invariant) regularized 
effective action in a closed 
form as well (or at least can be given an integral representation). This has been profitably used to 
calculate the imaginary part of the effective action for fermions interacting with a constant 
background electromagnetic field which describes the decay rate of the vacuum \cite{schwinger}. 
However, solving the dynamical equations in \eqref{dynamicaleqn} is, in general, not easy when 
interactions are present. When the dynamical equations cannot be solved in a closed form, the 
method due to Schwinger leads to a perturbative determination of the effective action.

In the past couple of decades, there have been several attempts \cite{generalization,frenkel} to 
generalize the method due to Schwinger to finite temperature \cite{temp,das} and to determine the 
imaginary part of the effective action leading to conflicting results \cite{generalization}. In 
\cite{dasfrenkel} we have 
presented an alternative method for determining finite temperature effective actions for fermions 
interacting with an arbitrary background field. We believe that since the amplitudes at finite 
temperature are ultraviolet finite unlike those at zero temperature, it is not necessary to generalize 
the method due to Schwinger to finite temperature. After all, the proper time method was designed 
to provide a (gauge invariant) regularization which is not necessary at finite temperature. Therefore, 
we have proposed \cite{dasfrenkel} a direct method for evaluating finite temperature effective 
actions based mainly on the general properties of systems at finite temperature. In this paper we 
give a detailed description of this method along with various other aspects not discussed in 
\cite{dasfrenkel}.

As we have emphasized in \cite{dasfrenkel}, we believe that the real time 
formalism \cite{das} (we use the closed time path formalism due to Schwinger \cite{schwinger1}) is 
more suited for this purpose. We note that, in general,  the imaginary time formalism (the Matsubara 
formalism \cite{matsubara})  leads naturally to retarded and advanced amplitudes, but the Feynman 
(time ordered) amplitudes (beyond the two point function) cannot be consistently generated in this 
formalism \cite{evans}. (We emphasize that this is true for space-time dimensions $d\geq 2$. In 
$0+1$ dimension the Feynman and the retarded amplitudes coincide and, therefore, the imaginary 
time formalism naturally leads to Feynman amplitudes.) On the other hand, the effective action that 
we are interested in is precisely the one that generates Feynman amplitudes. In contrast to the 
imaginary time formalism, the effective action, when evaluated properly in the real time formalism, 
leads naturally not only to the Feynman amplitudes, but also to the retarded and the advanced 
amplitudes as we will show in examples. Furthermore, as we have emphasized earlier in 
\cite{das,tor}, the real time calculations can be carried out quite easily in the mixed space where the 
spatial coordinates have been Fourier transformed as we will describe in the following examples.

The present paper is organized as follows. In section {\bf II} we recapitulate our proposal 
\cite{dasfrenkel} for evaluating effective 
actions at finite temperature. In section {\bf III} we apply the method to evaluating the complete 
effective action at finite temperature for the $0+1$ dimensional QED. From the structure of this 
effective action, we show that all the temperature dependent retarded (advanced) amplitudes 
vanish. 
The temperature dependent effective action for Schwinger model, the $1+1$ dimensional massless 
QED, is discussed in detail in section {\bf IV} where we show that the temperature dependent 
retarded (advanced) amplitudes vanish in this theory as well. We present our conclusions and 
summarize future directions in section {\bf V}.

\section{Proposal} 

From the definition of the effective action \eqref{effaction} for a system of massive fermions 
interacting with an arbitrary background, it is straightforward to obtain   
\begin{equation}
\frac{\partial\Gamma_{\rm eff}}{\partial m} = \int dt d\mathbf{x}\ S (t,\mathbf{x}; t, \mathbf{x}),
\label{propagator0}
\end{equation}
where $S (t,\mathbf{x};t', \mathbf{x}')$ denotes the complete Feynman propagator for the fermion 
(including the factor $i$) in the presence of the background field. However, keeping in mind that the 
fermion may not always have a mass (say, for example, in the Schwinger model \cite{schwinger2}), 
we use alternatively the fact that the variation of the effective action with respect to the background 
field leads to the generalized fermion ``propagator" at coincident points (even though we use the 
same symbol as in \eqref{propagator0}, the exact meaning of $S$ below depends on the nature of 
the background field as we explain),
\begin{equation}
\frac{\delta \Gamma_{\rm eff}}{\delta A (t,\mathbf{x})} =  g S (t, \mathbf{x}; t, \mathbf{x}),
\label{propagator1}
\end{equation}
where  we are suppressing the Lorentz structure of the background field as well as
that of the generalized ``propagator". We note that for a scalar
background, $S$ in \eqref{propagator1} indeed denotes the complete
fermion propagator of the interacting theory at coincident
coordinates. On the other hand, for a gauge field background, the
right hand side in \eqref{propagator1} determines the current density of the theory which is related 
to the complete fermion propagator of the theory through a Dirac trace involving the Dirac matrix. In 
either case, we note that it is the fermion propagator that is relevant in \eqref{propagator1} for the 
evaluation of the effective action. We note that in the mixed space (where the spatial coordinates 
$\mathbf{x}$ have been Fourier transformed), we can write \eqref{propagator1} as
\begin{equation}
\frac{\delta\Gamma_{\rm eff}}{\delta A (t, -\mathbf{p})} = g S (t,t; \mathbf{p}).\label{propagator}
\end{equation} 

Since the effective action is so intimately connected with the fermion propagator, our proposal is to 
determine the complete fermion propagator at finite temperature directly such that 
\begin{enumerate}
\item[(i)] it satisfies the appropriate equations for the complete propagator of the theory, 
\item[(ii)] it satisfies the necessary symmetry properties of the theory such as the Ward identity, 
\item[(iii)] and most importantly, it satisfies the anti-periodicity property associated with a finite 
temperature fermion propagator \cite{das}. 
\end{enumerate}
In fact, it is the third requirement that is quite important in a direct determination of the propagator. 
We note that this last condition is missing at zero temperature which makes it difficult to determine 
the complete propagator (independent of the problem of divergence). When the theory is free of 
ultraviolet divergence (so that it does not need a regularization at zero temperature), this propagator 
will be the exact fermion propagator of the theory and would lead to the complete effective action 
including the correct zero temperature part. On the other hand, if the theory needs to be regularized 
at zero temperature, this propagator will not yield the correct zero temperature effective action. 
However, we note that our interest is in the finite temperature part of the effective action which does 
not need to be regularized (it is not ultraviolet divergent) and will be determined correctly in this
 approach. We illustrate the method with two examples.

\section{$0+1$ dimensional QED}

Let us consider the $0+1$ dimensional QED described by the Lagrangian
\begin{equation}
L = \overline{\psi} (t) (i\partial_{t} - m - eA (t)) \psi (t),\label{0plus1L}
\end{equation}
where the fermion mass can be thought of as a chemical potential and in $0+1$ dimension, the 
gauge potential has only a single component (which we suppress). There are no Dirac matrices 
and as a result the gauge background behaves like a scalar background. This is a simple 
model which has been studied exhaustively \cite{0+1, dunne} in connection with large gauge 
invariance \cite{babu} at finite temperature, but it is also quite useful in clarifying various concepts
involved in our proposal before we generalize it to higher dimensions. 

\subsection{Finite temperature propagator}

\begin{figure}[ht!]
\begin{center}
\includegraphics{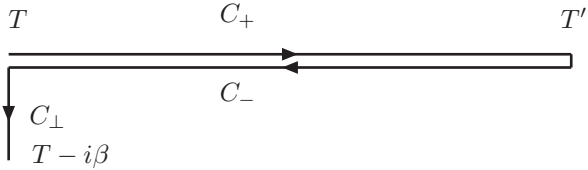}
\end{center}
\caption{The closed time path contour in the complex $t$ plane. Here
  $T\rightarrow -\infty$, while $T'\rightarrow \infty$ and $\beta$
  denotes the inverse temperature (in units of the Boltzmann constant
  $k$) \cite{das}.}
\label{1}
\end{figure}

As we noted earlier, we use the closed time path formalism 
where the path in the complex time plane has the form shown in Fig. \ref{1}. In the closed time path 
formalism (in any real time formalism) \cite{das}, the degrees of freedom need to be doubled and we 
denote the background fields on the $C_{\pm}$ branches of the contour as $A_{\pm} (t)$ 
respectively. The two branches labelled by $C_{\pm}$ lead to the
 doubling of the degrees of freedom while the branch $C_{\perp}$ along the imaginary axis 
decouples from any physical amplitude. 

Since $t$ is the only coordinate on which field variables depend in this theory, there is 
no need for a mixed space propagator. We note that the complete fermion propagator of the theory 
(ordered along the contour in Fig. \ref{1}) satisfies the equations
\begin{eqnarray}
& & (i\partial_{t} - m - e A_{c} (t)) S_{c} (t,t') = i\delta_{c} (t-t'),\nonumber\\
& & S_{c}(t,t') (i\overleftarrow{\partial}_{\!\!t'} + m + e A_{c} (t')) = - i\delta_{c} (t-t'),\label{0plus1eqn}
\end{eqnarray}
where the subscript ``$c$" characterizes a function on the contour. On the contour, the step function 
is defined naturally as \cite{das}  
\begin{equation}
\theta_{c} (t-t') = \left\{\begin{array}{ll}
\theta (t-t') & {\rm if\ both}\  t,t'\in C_{+},\\
\theta (t'-t) & {\rm if\ both}\ t,t' \in C_{-}\, {\rm or}\ \in C_{\perp},\\
1 & {\rm if}\ t\in C_{-}\ ({\rm or}\,\in C_{\perp}) {\rm and}\ t'\in C_{+},\\
0 & {\rm if}\ t\in C_{+} {\rm and}\ t'\in C_{-}\ ({\rm or}\,\in C_{\perp}),
\end{array}\right.\label{thetac}
\end{equation}
and the delta function on the contour is defined in the standard manner as
\begin{equation}
\delta_{c} (t-t') = \partial_{t} \theta_{c} (t-t').
\end{equation}

Equations \eqref{0plus1eqn} can be solved exactly subject to our three requirements in the 
following manner. First we note that a general solution of \eqref{0plus1eqn} can be written (in the 
contour ordered form) as
\begin{eqnarray}
S_{c} (t,t') & = &  \Big(\theta_{c} (t-t')\, C - \theta_{c} (t'-t)\, D\Big)\nonumber\\
& & \quad\times\,e^{-im(t-t') -ie\int\limits_{t'}^{t} dt''_{c}\,A_{c} (t''_{c})},\label{0plus1generalS}
\end{eqnarray}
where  $C$ and $D$ are independent of $t,t'$ satisfying 
\begin{equation}
C + D = 1,\label{0plus1condn1}
\end{equation}
in order for \eqref{0plus1generalS} to satisfy \eqref{0plus1eqn}. If we now impose the 
anti-periodicity condition
\begin{equation}
S_{c} (-\infty, t') = - S_{c} (-\infty-i\beta,t'),\label{0plus1antiperiodicity}
\end{equation}
where $\beta$ is the inverse of temperature in units of the Boltzmann constant ($k$) we determine
\begin{equation}
D = C\,e^{-\beta m - ie\int\limits_{-\infty}^{-\infty-i\beta} dt'_{c}\,A_{c} (t'_{c})} = C\,e^{-\beta m -ie 
(a_{+}-a_{-})}.\label{0plus1condn2}
\end{equation}
Here we have defined
\begin{equation}
a_{\pm} = \int\limits_{-\infty}^{\infty} d t\, A_{\pm} (t),\label{0plus1a}
\end{equation}
and have used the fact that the vertical branch decouples from amplitudes.

Together with \eqref{0plus1condn1} the relation in \eqref{0plus1condn2} determines
\begin{equation}
C = 1 - n_{\rm \sc F} \big(m + \frac{ie}{\beta} (a_{+}-a_{-})\big),\quad D = n_{\rm\sc F} \big(m + 
\frac{ie}{\beta} (a_{+}-a_{-})\big),\label{0plus1constants}
\end{equation}
where $n_{\rm F}$ denotes the Fermi distribution function and leads to the contour ordered
propagator of the form
\begin{eqnarray}
S_{c} (t,t') & = & e^{-imt-ie\int d \bar{t}\, \theta_{c} (t-\bar{t}) A_{c} (\bar{t})}\nonumber\\
& & \times\left(\theta_{c} (t-t') - n_{\rm F} (m+\frac{ie}{\beta} (a_{+} - a_{-}))\right)\nonumber\\
& & \times e^{imt'+ie \int d \bar{t}\, \theta_{c}(t'-\bar{t}) A_{c} (\bar{t})}.
\label{0plus1propagator}
\end{eqnarray}
In fact, this propagator satisfies anti-periodicity in both its arguments, namely, 
\eqref{0plus1antiperiodicity} 
as well as the condition
\begin{equation}
S_{c} (t,-\infty) = - S_{c} (t,-\infty-i\beta).\label{0plus1antiperiodicity1}
\end{equation}
Although the phases in \eqref{0plus1propagator} can be combined to write them in a simpler form 
as in \eqref{0plus1generalS} in 
this case, we have chosen to write them in this suggestive form which generalizes naturally to 
higher dimensions where the propagator will carry spinor indices. When $t,t'$ are restricted to the 
appropriate branches of the contour, \eqref{0plus1propagator} determines all the components of the 
full $2\times 2$ matrix propagator of the theory,
\begin{eqnarray}
S_{++} (t,t') & = & \!\!\Big(\theta(t-t') - D\Big)e^{-im(t-t') -ie\int\limits_{t'}^{t} dt'' A_{+}(t'')},\nonumber\\
S_{+-} (t,t') & = & \!\!- D e^{-im(t-t') + ie(\int\limits_{t}^{\infty}dt'' A_{+} (t'') + \int
\limits_{t'}^{\infty}dt'' A_{-}(t''))},\nonumber\\
S_{-+} (t,t') & = &  \!\!Ce^{-im(t-t') -ie(\int\limits_{t}^{\infty} dt'' A_{-}(t'') + \int\limits_{t'}^{\infty} dt'' A_{+}
(t''))},\nonumber\\
S_{--} (t,t') & = & \!\!-\Big(\theta(t-t') -C\Big)e^{-im(t-t') + ie\int\limits_{t'}^{t}dt'' A_{-}(t'')},\nonumber\\
& &
\end{eqnarray}
where the constants $C,D$ are defined in \eqref{0plus1constants}.

\subsection{Lippmann-Schwinger equation}

In addition to satisfying the anti-periodicity conditions, it can be checked that the propagator in 
\eqref{0plus1propagator} satisfies the required Ward identity.  Furthermore, it also satisfies the 
Lippmann-Schwinger equation (the perturbation expansion) \cite{lippman} for the 
propagator which can be seen as follows. For simplicity of discussion, let us consider the 
propagator only  in the $C_{+}$ branch of the contour in the complex $t$-plane and factor out the 
exponential factor $e^{-im (t-t')}$ which trivially factors out in a product. In this case, we can 
write (suppressing the thermal index $+$)
\begin{equation}
S (t,t') =  e^{-ie\Phi (t)} \Big(\theta (t-t') - D\Big)e^{ie\Phi (t')},\label{ls1}
\end{equation}
where we have defined ($D$ is defined in \eqref{0plus1constants}) 
\begin{equation}
\Phi (t) = \int dt''\,\theta(t-t'') A (t''),\label{ls2}
\end{equation}
with
\begin{equation}
 \Phi (\infty) = a = \int\limits_{-\infty}^{\infty} dt\,A(t),\quad \Phi (-
\infty) = 0.\label{ls2'}
\end{equation}
The propagator, in the absence of interactions, is given by
\begin{equation}
S_{0} (t,t') = \theta (t-t') - n_{\rm\sc F} (m).\label{ls3}
\end{equation}
The Lippmann-Schwinger equation can be written as
\begin{equation}
S (t,t') = \big(1 + ie S_{0} A\big)^{-1} S_{0}\,(t,t'),\label{ls4}
\end{equation}
which can also be expressed as
\begin{eqnarray}
& & \big(1+ie S_{0} A\big) S (t,t') = S_{0} (t,t'),\nonumber\\
{\rm or,}& & S(t,t') - S_{0} (t,t') + ie S_{0} A S (t,t') = 0.\label{ls5}
\end{eqnarray}

Using the definition \eqref{ls2} we note that we can write
\begin{equation}
A (t) =  \frac{d\Phi (t)}{dt},\label{ls6}
\end{equation}
which leads to 
\begin{align}
ie S_{0}A S ( t,t') & = - \int dt'' S_{0} (t,t'') \frac{de^{-ie\Phi (t'')}}{dt''}\nonumber\\
& \quad \times \Big(\theta (t''-t') - D\Big) e^{ie\Phi (t')}.\label{ls7}
\end{align}
Integrating the right hand side of \eqref{ls7} (by parts) and using \eqref{ls2'} as well as identities 
associated with Fermi distribution functions we obtain
\begin{equation}
ie S_{0} A S (t,t') = - S (t,t') + S_{0} (t,t').\label{ls8}
\end{equation}
Using this in \eqref{ls5} we conclude that the propagator \eqref{ls1} satisfies the 
Lippmann-Schwinger equation to all orders. This argument can be carried over to show that the 
complete $2\times 2$ matrix propagator satisfies the Lippmann-Schwinger equation. In other words, 
the complete propagator \eqref{0plus1propagator} satisfies exactly the perturbative expansion (for 
the propagator) to all orders in the interaction at finite temperature.

\subsection{Effective action}

The $0+1$ dimensional theory is free from ultraviolet divergences and, therefore, 
\eqref{0plus1propagator}  
represents the complete fermion propagator of the theory in the presence of a background gauge 
field. We can now take the coincident limit $(t=t')$ in \eqref{0plus1propagator} to obtain
\begin{equation}
S_{c} (t,t) = \frac{1}{2}\Big(1 - 2 n_{\rm\sc F} (m+\frac{ie}{\beta} (a_{+}-a_{-}))\Big).\label{0plus1ea1}
\end{equation}
It is interesting to note that in the coincident limit the propagator is independent of the time 
coordinate. This is, in fact, a consequence of the Ward identity of the theory (see \cite{dunne}). 

Using \eqref{0plus1ea1} we can integrate \eqref{propagator1} or \eqref{propagator} (note that there 
is no momentum in this $0+1$ dimensional example) to obtain the 
normalized effective action of the theory which has the form
\begin{eqnarray}
\Gamma_{\rm eff} [a_{+},a_{-}] & = & - i \ln \left[\cos \frac{e(a_{+}-a_{-})}{2}\right.\nonumber\\
& & \qquad\left. + i \tanh \frac{\beta m}{2}\sin \frac{e(a_{+}-a_{-})}{2}\right].\label{0plus1effaction}
\end{eqnarray}
This is the complete effective action of the theory which reduces to the well studied action 
\cite{0+1,dunne} on $C_{+}$ when we set $a_{-}=0$. However, being the complete effective action, 
\eqref{0plus1effaction} contains all the information about retarded, advanced and other amplitudes 
as well. 

For example, from the structure of the complete effective action in \eqref{0plus1effaction} we can 
now show that all the retarded (advanced) amplitudes in this theory vanish at finite temperature. Let 
us recall that the retarded $n$-point amplitude in a theory can be expressed as \cite{retarded} 
\begin{eqnarray}
\Gamma^{(n)}_{R} & = & \Gamma_{++\cdots +} + \Gamma_{+-+\cdots +} + \Gamma_{++-+\cdots +}
\cdots \nonumber\\
& & + \cdots + \Gamma_{+--\cdots -+} + \Gamma_{+--\cdots --}.\label{retardedamp}
\end{eqnarray}
Namely, it is the sum of all amplitudes where the first index is held fixed to be $+$ and all the $
(n-1)$ other thermal indices are permuted over the two values $\pm$. Let us note from 
\eqref{0plus1effaction}  that since $\Gamma_{\rm eff} = 
\Gamma_{\rm eff} [a_{+}-a_{-}]$, it now follows from \eqref{retardedamp}  that the retarded $n$-point 
amplitude of the theory  can be written as (since the amplitudes are time independent taking 
derivative with respect to the background field $A_{\pm} (t)$ is equivalent to taking derivative with 
respect to $a_{\pm}$) 
\begin{eqnarray}
\Gamma_{R}^{(n)} & = & \sum_{m=0}^{n-1} {}^{n-1}\!C_{m}\,\frac{d^{n-1-m}}
{da_{+}^{n-1-m}} \frac{d^{m}}{d a_{-}^{m}}\,\frac{d
\Gamma_{\rm eff}[a_{+}-a_{-}]}{da_{+}}\bigg|\nonumber\\
& = &\!\!\sum_{m=0}^{n-1}\!{}^{n-1}\!C_{m} \frac{d^{n-1-m}}
{d a_{+}^{n-1-m}} \!\!\left(-\frac{d }{d  a_{+}}\right)^{m}\,
\frac{d 
\Gamma_{\rm eff}[a_{+}-a_{-}]}{d a_{+}}\bigg|\nonumber\\
& = & (1-1)^{n-1}\, \frac{d ^{n}\Gamma_{\rm eff}[a_{+}-a_{-}]}{d  a_{+}^{n}}\bigg|  = 
0, \quad n\geq 2,\label{vanishingR}
\end{eqnarray}
where the restriction stands for setting all the background fields to zero. Therefore, all the retarded 
(advanced) amplitudes vanish in this theory. (The one point amplitude is by definition a Feynman 
amplitude.) 

\section{Schwinger model}

After the derivation of the complete effective action in the $0+1$ dimensional theory, let us next 
consider the fermion sector of the Schwinger model \cite{schwinger2} or massless QED in $1+1$ 
dimensions described by the Lagrangian density
\begin{equation}
{\cal L} = \overline{\psi} (t,x) \gamma^{\mu} (i\partial_{\mu} - eA_{\mu} (t,x)) \psi  (t,x).
\label{schwingerL}
\end{equation}
At zero temperature, this model is soluble and describes free massive photons. The effective action 
for this model (for an arbitrary gauge background) has also been studied perturbatively at finite 
temperature \cite{adilson} even in the presence of a chemical potential \cite{silvana}. Here we will 
derive the closed form expression of the finite temperature effective action following our method. We 
note here that the two point function in the Schwinger model needs to be regularized at zero 
temperature (this is the only non-vanishing amplitude in the Schwinger model at zero temperature)  
and, consequently, the zero temperature part of the effective action following from our 
propagator will not coincide with the regularized zero temperature effective action. However, our 
interest is in the finite temperature part of the effective action which is free from ultraviolet 
divergences. For completeness we note that the simple point-splitting regularization of the fermion 
propagator is sufficient to regularize the theory and can be carried out even in our method. 
However, we will not do this here since our main interest is in the finite temperature part of the 
effective action (which does not need a regularization).

The theory \eqref{schwingerL} is best studied in the natural basis of right handed and left handed 
fermion fields (although everything that we say can be carried out covariantly as well as in the 
presence of a chemical potential). Defining \cite{dasbook,glf}
\begin{eqnarray}
\psi_{\rm R} & = &  \frac{1}{2} (\mathbbm{1}+\gamma_{5})\psi, \quad \psi_{\rm L} = \frac{1}{2} 
(\mathbbm{1} - \gamma_{5}) \psi,\nonumber\\
x^{\pm} & = & \frac{x^{0}\pm x^{1}}{2}, \quad p_{\pm} = p_{0}\pm p_{1},\quad \partial_{\pm} = 
\partial_{0}\pm \partial_{1},\nonumber\\ 
A_{\pm} & = & A_{0} \pm A_{1},\label{light-conevariables}
\end{eqnarray}
the Lagrangian density \eqref{schwingerL} naturally decomposes into two decoupled sectors 
described by 
\begin{equation}
{\cal L} = \psi_{\rm R}^{\dagger} (i\partial_{+} - eA_{+})\psi_{\rm R} + \psi_{\rm L}^{\dagger} (i
\partial_{-} - eA_{-})\psi_{\rm L},\label{rlsectors}
\end{equation}
where $\psi_{\rm R}, \psi_{\rm L}$ denote only the component spinor fields (no spinor index left any 
more). While the zero temperature regularization mixes the two sectors through the two point 
function (anomaly), at finite temperature we do not have divergences and, therefore,  we do not 
expect the two sectors to mix. Therefore, we can study the finite temperature effective action in each 
of the two sectors separately.

\subsection{Propagator}

Let us consider the theory only in the sector of the right handed
fermions in \eqref{rlsectors}. This is very much like the $0+1$
dimensional theory. However, there is one essential difference which
makes the derivation much more difficult, namely, the field variables
depend on two coordinates $(t,x)$ or equivalently on
$(x^{+},x^{-})$. We would like to emphasize here that although we use
the light-cone coordinates for simplicity, the theory is still
quantized on the equal-time surface and the propagator is defined
through the time ordered Green's function (namely, we do not use the
statistical mechanics of the light-front \cite{lightfront}). As we
mentioned earlier, the finite temperature derivations become a lot
simpler in the mixed space. Thus, Fourier transforming the $x^{-}$
coordinate, the action for the right handed fermions takes the form
(the conjugate variables to $x^{-}$ should be written as $p_{-},k_{-}$, which we write as $p,k$ for 
simplicity)
\begin{eqnarray}
S_{\rm R} & = & 2 \int d x^{+} \frac{d p}{2\pi}\,\psi_{\rm R}^{\dagger} (x^{+},-p)
\bigg[i\partial_{+} \psi_{\rm R} (x^{+},p)\nonumber\\
 & & \quad  -e\int \frac{d k}{2\pi}\, A_{+} (x^{+},p-k) \psi_{\rm R} (x^{+},k)\bigg].\label{rtreeaction}
\end{eqnarray}

As a result, we recognize that the equations for the propagator will involve a convolution.They are 
best described by introducing the following operator notations for the propagator as well as the 
gauge potential
\begin{eqnarray}
S (x^{+},x'^{+};p,k) & = & \langle p|\hat{S} (x^{+},x'^{+})|k\rangle,
\nonumber\\ 
A_{+} (x^{+},p-k) & =  & \langle p|\hat{A}_{+} (x^{+})|k\rangle.\label{operators}
\end{eqnarray}
For example, the propagator equation following from \eqref{rtreeaction}
\begin{align}
&i\partial_{+} S (x^{+},x'^{+};p,k) - e \!\int \!\frac{dq}{2\pi}\,A_{+} (x^{+}, p-q) S (x^{+},x'^{+};q,k)
\nonumber\\
& \qquad = \frac{i}{2} \delta(x^{+}-x'^{+}) 2\pi\delta(p-k),
\end{align}
can be written in the compact form
\begin{equation}
(i\partial_{+} - e \hat{A}_{+} (x^{+})) \hat{S} (x^{+},x^{-}) = \frac{i}{2} \delta (x^{+}-
x'^{+}),
\end{equation}
with the operator notation. Here we have assumed the normalization of the momentum states to be
\begin{equation}
\langle p|k\rangle = 2\pi \delta (p-k).
\end{equation}
With the operator notation in \eqref{operators} the equations for the propagator ordered along the 
contour takes the (operator) forms (see also \eqref{0plus1eqn})
\begin{eqnarray}
&&\!\!\!(i\partial_{+} - e \hat{A}_{c} (x^{+})) \hat{S}_{c} (x^{+},x'^{+}) = \frac{i}{2}\,\delta_{c} (x^{+}-
x'^{+}),\nonumber\\
&&\!\!\!\!\!\!\!\!\! \hat{S}_{c} (x^{+},x'^{+}) (i\overleftarrow{\partial}'_{\!\!+} + e \hat{A}_{c} (x'^{+})) = - 
\frac{i}{2}\, \delta_{c} (x^{+}-x'^{+}).\label{1plus1eqn}
\end{eqnarray}
We note from \eqref{propagator} and \eqref{rtreeaction} that, in the present case,  we can identify
\begin{equation}
S_{c} (x^{+},x'^{+};p) = \int \frac{\mathrm{d}k}{2\pi}\, \langle k+p|\hat{S}_{c} (x^{+},x'^{+})|k\rangle.
\end{equation} 

The general solution of \eqref{1plus1eqn} can be written in the form
\begin{eqnarray}
\hat{S}_{c} (x^{+},x'^{+})  & = & \frac{1}{4}\, e^{-ie \int d \bar{x}^{+}\,\theta_{c} (x^{+}-
\bar{x}^{+}) \hat{A}_{+c} (\bar{x}^{+})}\nonumber\\
& & \times\, \big({\rm sgn}_{c} (x^{+}-x'^{+}) + \hat{\cal O}\big)\nonumber\\
& & \times\, e^{ie \int d \bar{x}^{+}\,\theta_{c} (x'^{+}-\bar{x}^{+}) \hat{A}_{+c} (\bar{x}^{+})},
\end{eqnarray}
where
\begin{equation}
{\rm sgn}_{c} (x^{+}-x'^{+}) = \theta_{c} (x^{+}-x'^{+}) - \theta_{c}(x'^{+}-x^{+}),
\end{equation}
and $\hat{\cal O}$ which contains all the nontrivial information about interactions and temperature is 
independent of the coordinates $x^{+},x'^{+}$.  Since $x^{-},x'^{-}$ have been Fourier transformed, 
the anti-periodicity can be imposed only at the level of matrix elements in the mixed space by  
requiring
\begin{equation}
\langle p|\hat{S} (\frac{-\infty + x}{2},x'^{+})|k\rangle = - e^{-\frac{\beta p}{2}} \langle p|\hat{S} (\frac{-
\infty-i\beta + x}{2},x'^{+})|k\rangle,\label{1plus1antiperiodicity}
\end{equation}
The factor $e^{-\frac{\beta p}{2}}$ in \eqref{1plus1antiperiodicity} arises basically from the 
exponential factor in the Fourier transform. The requirement of anti-periodicity 
\eqref{1plus1antiperiodicity} determines
\begin{align}
\hat{\cal O} & = 1 - 2(\hat{\cal O}_{+} + 1)^{-1},\nonumber\\
 \hat{\cal O}_{+} & = e^{\frac{ie (\hat{a}_{+ (+)}-\hat{a}_{+ (-)})}{2}}\,e^{\frac{\beta \hat{K}}{2}}
\,e^{\frac{ie (\hat{a}_{+(+)}-\hat{a}_{+(-)})}{2}},\label{operatorO}
\end{align}
where $\hat{K}$ denotes the momentum operator satisfying
\begin{equation}
\hat{K} |p\rangle = p |p\rangle,
\end{equation}
and ($(\pm)$ with the parenthesis denote the thermal 
indices while $+$ without the parenthesis represents the light-cone component of the background 
field)
 \begin{equation}
 \hat{a}_{+ (\pm)} = \int\limits_{-\infty}^{\infty} d x^{+}\,\hat{A}_{+(\pm)} (x^{+}).
 \end{equation}

Therefore, the contour ordered propagator satisfying the Ward identity as well as the appropriate 
anti-periodicity condition has the form
\begin{eqnarray}
\hat{S}_{c} (x^{+},x'^{+})  & = & \frac{1}{4}\, e^{-ie \int d \bar{x}^{+}\,\theta_{c} (x^{+}-
\bar{x}^{+}) \hat{A}_{+c} (\bar{x}^{+})}\nonumber\\
& & \times \big({\rm sgn}_{c} (x^{+}-x'^{+}) + 1 - 2(\hat{\cal O}_{+} + 1)^{-1}\big)\nonumber\\
& & \times e^{ie \int d \bar{x}^{+}\,\theta_{c} (x'^{+}-\bar{x}^{+}) \hat{A}_{+c} (\bar{x}^{+})},
\label{1plus1propagator}
 \end{eqnarray}
It can be checked that this propagator also satisfies the anti-periodicity condition on the second 
variable, namely,
\begin{equation}
\langle p|\hat{S} (x^{+},\frac{-\infty + x'}{2})|k\rangle = - e^{\frac{\beta k}{2}} \langle p|\hat{S} (x^{+},
\frac{-\infty-i\beta + x'}{2})|k\rangle.\label{1plus1antiperiodicity1}
\end{equation}
Therefore, it satisfies all the requirements of our proposal.

\subsection{Lippmann-Schwinger equation}

As in the $0+1$ dimension, we can show that the complete propagator \eqref{1plus1propagator} 
satisfies the Lippmann-Schwinger equation. Once again, for simplicity, we will restrict ourselves to 
the propagator on the $C_{+}$ branch (and we suppress the thermal index $+$). Let us note from 
\eqref{1plus1propagator} that in the absence of interactions the propagator at finite temperature can 
be written as 
\begin{align}
S_{0} (x^{+},x'^{+}) & = \frac{1}{4}\Big({\rm sgn} (x^{+}-x'^{+}) + \hat{O}_{0}\Big)\nonumber\\
& = \frac{1}{4}\Big({\rm sgn} (x^{+}-x'^{+}) + 1 - 2 n_{\sc\rm F} \big(\frac{\hat{K}}{2}\big)\Big). 
\end{align}
Furthermore, we define
\begin{equation}
\hat{\Phi} (x^{+}) = \int d\bar{x}^{+}\,\theta (x^{+}-\bar{x}^{+})\,\hat{A}_{+} (\bar{x}^{+}),
\label{1plus1phi}
\end{equation}
satisfying
\begin{equation}
\hat{\Phi} (\infty) = \hat{a}_{+} = \int dx^{+}\,A_{+} (x^{+}),\quad \hat{\Phi} (-\infty) = 0,
\end{equation}
so that the complete propagator at finite temperature \eqref{1plus1propagator} can be written as
\begin{equation}
S (x^{+},x'^{+}) = e^{-ie \hat{\Phi} (x^{+})}\,\frac{1}{4} \Big({\rm sgn} (x^{+}-x'^{+}) + \hat{O}\Big)e^{ie 
\hat{\Phi} (x'^{+})},\label{1plus1propagator1}
\end{equation}
where $\hat{O}$ is defined in \eqref{operatorO}.

In terms of these operators Lippmann-Schwinger equation can be written as (we point out here that 
the unconventional factor of $2$ in the interaction term is a consequence of the use of light-cone 
coordinates which can also be understood from the factor of $\frac{1}{2}$ in the equation 
\eqref{1plus1eqn})
\begin{equation}
S (x^{+},x'^{+}) = \big(1 + 2ie S_{0} A\big)^{-1} S_{0}\,(x^{+},x'^{+}),
\end{equation}
which also has the equivalent description
\begin{align}
& \big(1+2ie S_{0} A\big) S (x^{+},x'^{+}) = S_{0} (x^{+},x'^{+}),\nonumber\\
{\rm or,}\quad&  S(x^{+},x'^{+}) - S_{0} (x^{+},x'^{+}) + 2ie S_{0} A S (x^{+},x'^{+}) = 0.\label{1plus1ls}
\end{align}
Using the identity
\begin{equation}
\hat{A}_{+} (x^{+}) = \frac{d\hat{\Phi}(x^{+})}{dx^{+}},
\end{equation}
we note that we can write
\begin{align}
2ie S_{0} \hat{A}_{+} S (x^{+},x'^{+}) & = - 2\int dx''^{+}\,S_{0} (x^{+},x''^{+}) \frac{d e^{-ie\hat{\Phi} 
(x''^{+})}}{dx''^{+}}\nonumber\\
& \quad\times\,\frac{1}{4}\Big({\rm sgn} (x''^{+}-x'^{+}) + \hat{O}\Big)e^{ie\hat{\Phi} (x'^{+})}.
\end{align}
Integrating this by parts, as in the $0+1$ dimensional case, it is straightforward to show that 
\eqref{1plus1ls} holds, namely, the Lippmann-Schwinger equation holds so that the complete 
propagator agrees with its perturbative expansion to all orders. 

 \subsection{Effective action}

We note that the propagator \eqref{1plus1propagator} involves operators which do not commute in 
general. Nonetheless it can be checked that at the coincident point, $x^{+}=x'^{+}$ the two 
exponential factors cancel each other and the propagator becomes independent of $x^{+}$ much 
like in the $0+1$ dimension. Let us recall that the effective action can be obtained from the 
propagator at the coincident limit (see, for example, \eqref{propagator}). This coordinate 
independence is expected from the equations of 
motion \eqref{1plus1eqn} as well as from the Ward identity of the theory. However, since the 
cancellation is a bit more involved than in the $0+1$ dimension, we discuss this in some detail.

To show the cancellation of phases in \eqref{1plus1propagator} to all orders when $x^{+}=x'^{+}$, 
let us recall the definitions in \eqref{operators} and \eqref{1plus1phi} which lead to
\begin{align}
\langle k+p|\hat{\Phi} (x^{+})|k\rangle &= \Phi (x^{+},p)\nonumber\\
& = \int dx''^{+} \theta_{c} (x^{+}-x''^{+}) A_{+c} 
(x''^{+},p).\label{1plus1ls2}
\end{align}
With this, let us look at the contributions coming only from the phases in the propagator 
\eqref{1plus1propagator} or \eqref{1plus1propagator1} at any fixed order $N$ when $x^{+}=x'^{+}$, 
namely,
\begin{align}
S_{c} (x^{+};p)\big| & = S_{c} (x^{+},x^{+};p)\big|\nonumber\\
& = \int \frac{dk}{8\pi} \langle k+p|e^{-ie\hat{\Phi}(x^{+})}\hat{O} e^{ie\hat{\Phi} 
(x^{+})}|k\rangle\big|.\label{Norder}
\end{align}
Here the restriction stands for looking at only the $N$th order terms coming from the exponentials. 
Introducing sets of complete momentum states, we can write this as (we suppress the $x^{+}$ 
dependence in $\Phi(x^{+},p)$)
\begin{align}
S_{c} (x^{+};p)\big| & = \big(-ie\big)^{N} \sum_{m=0}^{N} {}^{N}\!C_{m} (-1)^{m}\nonumber\\
& \times\!\!\int \!\!\frac{dk}{8\pi}\frac{dp_{1}}{2\pi}\cdots \frac{dp_{N+1}}{2\pi}2\pi\delta(p-p_{1}-\cdots - 
p_{N+1})\nonumber\\
& \times \left(\prod_{i=1}^{m}\Phi (p_{i})\right)\left(\prod_{j=m+2}^{N+1}\Phi(p_{j})\right)\nonumber\\
&\times\langle k+p_{1}+\cdots +p_{m+1}|\hat{O}|k+p_{1}+\cdots + p_{m}\rangle,\label{Norder1}
\end{align}
where we understand that the first product is unity for $m=0$ while the second factor is unity for 
$m=N$. Redefining $k\rightarrow k-p_{1}-\cdots - p_{m}$ followed by the transformation 
$p_{1}\leftrightarrow p_{m+1}$, equation \eqref{Norder1} can be written as
\begin{align}
S_{c} (x^{+};p)\big| & \!\!= \!\!\int \frac{dk}{8\pi}\frac{dp_{1}}{2\pi}\cdots \frac{dp_{N+1}}{2\pi}2\pi
\delta(p-p_{1}-\cdots - p_{N+1})\nonumber\\
& \times \Phi(p_{2})\Phi(p_{3})\cdots \Phi(p_{N+1})\langle k+p_{1}|\hat{O}|k\rangle\nonumber\\
& \times (-ie)^{N} \sum_{m=0}^{N} {}^{N}\!C_{m} (-1)^{m}.\label{Norder2} 
\end{align}
The sum in the last line of \eqref{Norder2} simply vanishes because
\begin{equation}
\sum_{m=0}^{N} {}^{N}\!C_{m} (-1)^{m} = (1-1)^{N} = 0.
\end{equation}
This shows that the contributions coming from the exponentials at any order $N$ vanish when 
$x^{+}=x'^{+}$ so that the phases do not contribute to the propagator in the coincident limit in spite 
of the fact that these now involve non-commuting operators. As a result, the propagator 
\eqref{1plus1propagator} can be written in the coincident limit as
\begin{align}
S_{c} (x^{+};p) & = \int \frac{dk}{8\pi}\,\langle k+p|\hat{O}|k\rangle\notag\\
& = \int \frac{dk}{2\pi}\,\langle k+p|\frac{1}{4}( 1 - 2 (\hat{\cal O}_{+} + 1)^{-1})|k\rangle,
\label{coincidentlimit}
\end{align}
and since the operator $\hat{O}$ \eqref{operatorO} is independent of coordinates, it follows that the 
propagator is independent of the coordinate $x^{+}$ in the coincident limit
\begin{equation}
\partial_{+} S_{c}(x^{+};p) = 0.
\end{equation}
This is consistent with the Ward identity of the theory and also follows from the equations of motion 
\eqref{1plus1eqn}.

Once the coincident limit of the propagator is determined, we can integrate \eqref{propagator} to 
determine the normalized effective action in the right handed sector in the following manner.  We 
recall the definitions in \eqref{operatorO} which leads, for example, to
\begin{equation}
\frac{d\hat{\cal O}_{+}}{d\hat{a}_{+}} = ie\,\hat{\cal O}_{+}.\label{Oderivative}
\end{equation}
Let us define
\begin{equation}
\hat{F} = \frac{1}{2}\,\ln \hat{\cal O}_{+},\label{F}
\end{equation}
in terms of which we can write
\begin{align}
& \hat{\cal O}_{+}  = e^{2\hat{F}},\nonumber\\
& \hat{\cal O}  = 1 - 2 (\hat{\cal O}_{+} + 1)^{-1} = 1 - 2\big(e^{2\hat{F}} + 
1\big)^{-1}.\label{O}
\end{align}

From the definition in \eqref{F} as well as \eqref{Oderivative}, it follows that
\begin{equation}
\frac{d\hat{F}}{d\hat{a}_{+}} = \frac{1}{2\hat{\cal O}_{+}}\,\frac{d\hat{\cal O}_{+}}{d\hat{a}_{+}} = 
\frac{ie}{2}.
\end{equation}
Using this, we can now derive the well defined derivative 
\begin{align}
\frac{d\ln \cosh \hat{F}}{d\hat{a}_{+}} & = \frac{ie}{2}\big(e^{\hat{F}} + e^{-\hat{F}}
\big)^{-1}\big(e^{\hat{F}} - e^{-\hat{F}}\big)\nonumber\\
& = \frac{ie}{2} \big(1 - 2 (e^{2\hat{F}} + 1)^{-1}\big)\nonumber\\
&  = \frac{ie}{2}\big(1 - 2(\hat{\cal O}_{+} + 
1)^{-1}\big)= \frac{ie}{2}\,\hat{\cal O}.
\end{align}
Using these relations we can write the complete normalized effective action in the right handed 
sector as
\begin{align}
& \Gamma_{\rm R,\, eff}  =  - \frac{i}{2} \int \frac{dk}{2\pi}\,\langle k| \ln \cosh \hat{F} - 
\ln \cosh \frac{\beta\hat{K}}{4}|k\rangle\nonumber\\
& = - \frac{i}{2} \int \frac{dk}{2\pi}\,\langle k| \ln \cosh (\frac{1}{2} \ln \hat{\cal O}_{+}) - 
\ln \cosh \frac{\beta\hat{K}}{4}|k\rangle.\label{raction}
\end{align}
We note here that the perturbative expansion for the effective action is much easier to obtain from 
the propagator in \eqref{coincidentlimit}. 

The thermal part of this effective action has the right (delta function) structure that had already been 
observed in the 
perturbative calculation in the right handed sector \cite{adilson} which is a consequence of the Ward 
identity in the theory. (We remind the readers that we are not interested in the zero temperature part 
of this effective action which as we have argued would not correspond to the regularized action.)  
However, the expansion of this effective action on $C_{+}$ (namely, setting 
$A_{+(-)}=0$) does not  quite agree with the perturbative result order by order.  In fact, the difference 
already shows up in the quartic effective action. This is indeed very interesting and brings out the 
power of calculations in the mixed space which we have stressed repeatedly. The perturbative 
calculation \cite{adilson,silvana} which was carried out in momentum space misses out on a class of 
terms because of some subtlety that is not present in the mixed space. Namely, a class of terms had 
been set to zero in the perturbative calculation because of the identity
\begin{equation}
\frac{1}{p(p+q)} + \frac{1}{q(p+q)} - \frac{1}{pq} = 0.
\end{equation}
Although this identity is naively true, it does not hold when principal values are involved which is the 
case in the perturbative calculation. The correct identity in this case is 
\begin{equation}
\frac{1}{p(p+q)} + \frac{1}{q(p+q)} - \frac{1}{pq} = \pi^{2}\,\delta (p) \delta (q),\label{correction}
\end{equation}
and with this correction, the perturbative effective action in the right handed sector coincides exactly 
with the quartic effective action derived from \eqref{raction}. (It is worth pointing out here that the effective action 
\eqref{raction} is complete compared with the one proposed in \cite{dasfrenkel} which was not 
derived there.) We note here that in the leading 
order of the hard thermal loop approximation, when operators commute, this effective action 
coincides with that for the $0+1$ case for every value of the momentum with the identification 
$k=2m$. This is easily seen by comparing \eqref{coincidentlimit} in the commuting limit with 
\eqref{0plus1ea1}. (The extra factor of $\frac{1}{2}$ is associated with light-cone coordinates.)

The effective action for the left handed sector can similarly be derived and is given by 
\begin{align}
& \Gamma_{\rm L,\, eff}\nonumber\\
& = -\frac{i}{2} \int \frac{dk}{2\pi}\,\langle k|\ln \cosh (\frac{1}{2} \ln \hat{\cal O}_{-}) - 
\ln \cosh \frac{\beta\hat{K}}{4}|k\rangle,\label{laction}
\end{align}
where $k,p$ should be understood as the light-cone components conjugate to $x^{+}$
(which should be written as $k_{+},p_{+}$) and  
 we have identified
\begin{align}
& \hat{\cal O}_{-}  = e^{\frac{ie\hat{a}_{-}}{2}} e^{\frac{\beta\hat{K}}{2}} e^{\frac{ie\hat{a}_{-}}{2}},
\nonumber\\
& \hat{a}_{-}  = \hat{a}_{-(+)}-\hat{a}_{-(-)},\nonumber\\
& \hat{a}_{-(\pm)}  =  \int\limits_{-\infty}^{\infty} d x^{-}\,\hat{A}_{-(\pm)} (x^{-}).
\end{align}

Once again, the thermal part of this effective action has the right (delta function) structure as in the 
perturbative 
calculation and the thermal part agrees with the perturbative result order by order 
when restricted to $C_{+}$ (with the modification due to the subtlety discussed in \eqref{correction}). 
The finite temperature effective action for the  $1+1$ dimensional 
fermion interacting with an arbitrary Abelian gauge background can, therefore, be obtained from 
\begin{equation}
\Gamma_{\rm eff} = \Gamma_{\rm R, eff} + \Gamma_{\rm L, eff}\label{complete}
\end{equation}
and the thermal part of \eqref{complete} leads to the correct perturbative result order by order 
on the branch $C_{+}$ with the correction \eqref{correction}. Furthermore, since \eqref{complete} 
represents the complete 
effective action and since it is a functional of $(\hat{a}_{\pm (+)} - \hat{a}_{\pm (-)})$ (see 
\eqref{raction}-\eqref{laction}), it can be checked as in \eqref{vanishingR} that all the retarded 
(advanced) amplitudes vanish in this theory. This should be contrasted with the fact that this had 
been verified explicitly only up to the 4-point function in perturbation theory  \cite{retarded}.

\section{Summary} 

In summary, we have proposed an alternative method \cite{dasfrenkel} for determining effective 
actions at finite temperature for fermions interacting with an arbitrary background field. This is done 
by determining the complete fermion propagator (in the closed time path formalism) directly by 
using the anti-periodicity condition appropriate at finite temperature. We have illustrated in detail 
how our proposal works with the examples of the $0+1$ dimensional QED as well as the Schwinger 
model.  The next step in this direction would involve determining the effective action for massive 
QED in $1+1$ dimensions at finite temperature. This would lead directly to the finite temperature 
effective actions for the known solvable examples in four dimensional QED \cite{schwinger,soluble}

\bigskip

\noindent{\bf Acknowledgments}
\medskip

This work was supported in part  by US DOE Grant number DE-FG 02-91ER40685,  by CNPq and 
FAPESP (Brazil).


\begin{thebibliography}{10}

\bibitem{schwinger} J. Schwinger, Phys. Rev. {\bf 82}, 664 (1951).

\bibitem{generalization} W. Dittrich, Phys. Rev. {\bf D19}, 23 (1978); P. H. Cox, and W. S. Hellman, 
Ann. Phys. {\bf 154}, 211 (1984); M. Loewe and J. C. Rojas, Phys. Rev. {\bf D46}, 2689 (1992); P. 
Elmfors, D. Persson and B.-S. Skagerstam, Phys. Rev. Lett. {\bf 71}, 480 (1993); P. Elmfors and B.-S. 
Skagerstam, Phys. Lett. {\bf B348}, 141 (1995); A.K. Ganguly, J.C. Parikh, P.K. Kaw, Phys. Rev. {\bf C51}, 2091 (1995);  H. Gies, Phys. Rev. {\bf D60}, 105002 (1999); S. P. 
Gavrilov and D. Gitman, Phys. Rev. {\bf D78}, 045017 (2008); S. P. Kim,  H. K. Lee, Y. Yoon, Phys. Rev. {\bf D78}, 105013 (2008); S. P. Kim, H. K. Lee, Y. Yoon, arXiv:0910.3363 [hep-th]. 


\bibitem{frenkel} A. Das and J. Frenkel, Phys. Rev. {\bf D75}, 025021 (2007).

\bibitem{temp} J. Kapusta, {\em Finite Temperature Field Theory}, Cambridge University Press, 
Cambridge, England (1989); M. Le Bellac, {\em Thermal Field Theory}, Cambridge University Press, 
Cambridge, England (1996).

\bibitem{das} A. Das, {\em Finite Temperature Field Theory}, World Scientific, Singapore (1997).

\bibitem{dasfrenkel} A. Das and J. Frenkel, Phys. Lett. {\bf B680}, 195 (2009).

\bibitem{schwinger1} J. Schwinger, {\em Lecture Notes of Brandeis Summer Institute in Theoretical 
Physics} (1960); J. Schwinger, J. Math. Phys. {\bf 2}, 407 (1961); P. M. Bakshi and K. T. 
Mahanthappa, J. Math. Phys. {\bf 4}, 1 (1963); L. V. Keldysh, Sov. Phys. JETP {\bf 20}, 1018 (1965).

\bibitem{matsubara} T. Matsubara, Prog. Theor. Phys. {\bf 14}, 351 (1954).

\bibitem{evans} T. S. Evans, Nucl. Phys. {\bf B374}, 340 (1992).

\bibitem{tor}  F. T. Brandt, A. Das, O. Espinosa, J. Frenkel and S. Perez, Phys. Rev. {\bf D72}, 
085006 (2005); {\em ibid} {\bf D73}, 065010 (2006); {\em ibid} {\bf D73}, 067702 (2006).

\bibitem{schwinger2} J. Schwinger, Phys. Rev. {\bf 128}, 2425 (1962).

\bibitem{0+1} G. Dunne, K. Lee and C. Lu, Phys. Rev. Lett. {\bf 78}, 3434 (1997)

\bibitem{dunne} A. Das and G. Dunne, Phys. Rev. {\bf D57}, 5023 (1998); J. Barcelos-Neto and A. 
Das, Phys. Rev. {\bf D58}, 085022 (1998).

\bibitem{babu} K. S. Babu, A. Das and P. Panigrahi, Phys. Rev. {\bf D36}, 3725 (1987).

\bibitem{lippman} B. A. Lippmann and J. Schwinger, Phys. Rev. {\bf 79}, 469 (1950).

\bibitem{retarded} F. T. Brandt, A. Das, J. Frenkel and A. J. da Silva, Phys. Rev. {\bf D59}, 065004 
(1999); F. T. Brandt, A. Das and J. Frenkel, Phys. Rev. {\bf D60}, 105008 (1999).

\bibitem{adilson} A. Das and A. J. da Silva, Phys. Rev. {\bf D59}, 105011 (1999).

\bibitem{silvana}  S. Maciel and S. Perez, Phys. Rev. {\bf D78}, 065005 (2008).

\bibitem{dasbook}
For quantization of massless fermion fields, see, for example, {\em Lectures on Quantum Field 
Theory}, A. Das, World Scientific Publishing, Singapore (2008).

\bibitem{glf} For a general discussion of transformations between different coordinate systems, see, 
for example, A. Das and S. Perez, Phys. Rev. {\bf D70}, 065006 (2004).

\bibitem{lightfront} V. S. Alves, A. Das and S. Perez, Phys. Rev. {\bf D66}, 125008 (2002).

\bibitem{soluble} See, for example, G. Dunne, arXiv:hep-th/0406216; H. M. Fried and R. Woodard, 
Phys. Lett. {\bf B524}, 233 (2002).


\end{thebibliography}
\end{document}